\documentclass[superscriptaddress,book,showpacs]{revtex4}
\usepackage{graphics,graphicx}
\usepackage{amsfonts}
\input epsf

\newcommand{\dd}{\partial}
\newcommand{\be}{\begin{equation}}
\newcommand{\ee}{\end{equation}}
\newcommand{\bea}{\begin{eqnarray}}
\newcommand{\eea}{\end{eqnarray}}

\newcommand{\bra}{\langle}
\newcommand{\ket}{\rangle}
\newcommand{\lp}{\left(}
\newcommand{\rp}{\right)}

\newcommand{\up}{\uparrow}
\newcommand{\down}{\downarrow}

\begin{document}
\title{Teleportation from a Projection Operator Point of View}

\author{N. Erez}

\affiliation{Institute for Quantum Studies and Physics Dept., Texas
A\&M University, College Station, TX 77843-4242}

\vspace{.4cm}

\begin{abstract}
The process of quantum state teleportation is described from the
point of view of the properties of projections onto one-dimensional
subspaces. It is introduced as a generalization of the remote
preparation of a known state by use of an EPR pair. The discrete and
continuous cases are treated in a unified way. The conceptual and
calculational simplicity is pedagogically advantageous.
\end{abstract}


\maketitle

\section{Introduction}

Quantum state teleportation has become, since its introduction a
decade ago by Bennett et al.\cite{bennet}, one of the most important
(conceptual) applications of entanglement. This is illustrated by
the interesting fact that almost every recent paper on entanglement
begins by citing the use of that resource for teleportation.
Although the main result is given by a single equation which can be
verified with a few lines of elementary algebra (Equation 5 of
\cite{bennet}), it is subtle and raises many questions: What is the
precise role of entanglement, and how does it relate to other
manifestations of it? What role do the ($SU(2)$) group properties
play? That there was more than meets the eye to this apparently
irreducible derivation, was strikingly illustrated by Vaidman in his
interpretation of teleportation \cite{mytele} in terms of a class of
measurements introduced much earlier by Aharonov and
Albert\cite{AA}, which suggested to him a generalization to the
teleportation of states of a particle with a continuous degree of
freedom.

To me, teleportation is still easier calculated than understood. I
think the most natural motivation is the remote preparation of a
known spin state\cite{Pati} using a measurement on the other half of
an EPR-Bohm pair, which works half the time (and registers when it
fails). State teleportation can then be viewed as a generalization
to the remote preparation of an unknown state determined by the
state of an auxiliary particle. Interestingly, this problem is in
some respects simpler, and an efficient corrective step can be
implemented when the direct process fails (3/4 of the time in this
case). In view of the linearity of projection operators, the main
trick is to find the appropriate basis corresponding to the
(two-particle) measurements. The key is the basis element for which
no correction is needed. Continuous teleportation is then described
in analogous fashion.

The main results require nothing more than elementary linear algebra
(as is also true of the original teleportation paper). A few simple
group theoretic and information theoretic results are mentioned, but
only for motivation, or in comments.

\section{EPR-Bohm and remote state preparation}

The concept of entanglement had its origin in the famous EPR
paper\cite{EPR}. It is often more convenient to use Bohm's
version\cite{EPRB} of  a spin singlet state of two spatially
separated spin 1/2 particles. That is also our starting point.

The next two paragraphs will introduce remote state
preparation\cite{Pati} using Pauli spin algebra terminology and
properties (e.g., $ |\up_{\hat{n}} \ket $ will be used for the state
with spin projection $+\hbar/2$ along the $\hat{n}$ axis). This will
only serve as heuristic motivation. Once done, however, it will
prove profitable to rephrase the result in plain linear algebraic
language.

Suppose two experimenters, `Alice' and `Bob', share an EPR-Bohm
state, i.e., each has a spin 1/2 particle localized spatially in a
small space, but both sharing a total spin state of zero spin:
$\lp\frac{|\up\down\ket-|\down\up\ket }{\sqrt{2}}\rp_{12}
\psi(x_1-x_{\mathrm{Alice}})\psi(x_2-x_{\mathrm{Bob}})$. At this
point, the two particles each have an undetermined spin state. Alice
would like Bob's particle to be in the state $|\up_{\hat{n}} \ket$.
She can perform a measurement on her own particle with respect to
the $\hat{n}$ axis (i.e., the measurement basis is
$\left\{|\up_{\hat{n}}\ket, |\down_{\hat{n}} \ket \right\} $). The
two possible outcomes of the experiment can be expressed as follows:

\be \left\{
      \begin{array}{c}
        \lp |\up_{\hat{n}} \ket \bra \up_{\hat{n}} | \rp_1  \lp \frac{
|\up\down\ket-|\down\up\ket }{\sqrt{2}}\rp_{12} = \frac{1}{\sqrt{2}}
| \up_{\hat{n}} \ket_1| \down_{\hat{n}} \ket_2 \\
        \lp |\down_{\hat{n}} \ket \bra \down_{\hat{n}} | \rp_1  \lp \frac{
|\up\down\ket-|\down\up\ket }{\sqrt{2}}\rp_{12} = \frac{1}{\sqrt{2}}
| \down_{\hat{n}} \ket_1| \up_{\hat{n}} \ket_2 \\
        \end{array}
    \right. \label{stateprep1}
 \ee where we have used the symmetry property that $\frac{
|\up_{\hat{n}}\down_{\hat{n}}\ket-|\down_{\hat{n}}\up_{\hat{n}}\ket
}{\sqrt{2}}=\frac{
|\up_{\hat{n}'}\down_{\hat{n}'}\ket-|\down_{\hat{n}'}\up_{\hat{n}'}\ket
}{\sqrt{2}}$ for all $\hat{n},\hat{n}'$ (provided we had been
careful to define the phases of the $\{|\up_{\hat{n}}\ket\}$
consistently).

This means that half the time (i.e., with probability 1/2) Alice
measures $ | \down_{\hat{n}} \ket_1$ and she can tell Bob his
particle has the desired spin state. What about the other half?
Alice tells Bob his particle's spin is flipped. Can he fix it? The
operation  $ | \up_{\hat{n}} \ket \mapsto | \down_{\hat{n}} \ket
~~\forall \hat{n}$, is not unitary. There are unitary operations
that flip a $ | \up_{\hat{n}} \ket $ for a \emph{given} $\hat{n}$,
namely rotations about an axis perpendicular to $\hat{n}$. But if
Bob knows the direction $\hat{n}$, he might as well prepare $ |
\up_{\hat{n}} \ket $ locally.

Now let us rephrase this in simple linear algebraic terms. Fix an
orthonormal basis: $\left\{ |\up\ket,~~  | \down\ket \right\}$ where
'up' and 'down' will only be taken as convenient labels\footnote{The
notation is justified by the fact that there exists an axis with
respect to which this corresponds to the 'up' and 'down' states, for
appropriate choice of phases. We do not need this property.}. Thus,
$ | \up_{\hat{n}} \ket = \alpha^* | \up\ket +\beta^* | \down  \ket $
and $ | \down_{\hat{n}} \ket = \beta | \up\ket -\alpha | \down \ket
$ for some $\alpha,\beta\in\mathbb{C}$ (neglecting an
inconsequential common phase). We shall not use the symmetry of the
singlet state with respect to rotation, and it is convenient to
replace this 2-particle state shared by Alice and Bob by $\frac{
|\up\up\ket+|\down\down\ket }{\sqrt{2}}$ (in fact, we can just
relabel Alice's basis states so as to give the original state this
form with respect to the new basis). Equation \ref{stateprep1} now
reads:

\be \left\{
      \begin{array}{c}
        \left[ \lp \alpha^* | \up\ket +\beta^* | \down  \ket \rp \lp \alpha \bra \up |+
        \beta \bra \down | \rp \right]_1  \lp \frac{
|\up\up\ket+|\down\down\ket }{\sqrt{2}}\rp_{12} = \frac{1}{\sqrt{2}}
\lp \alpha^* | \up\ket +\beta^* | \down  \ket \rp_1 \lp \alpha | \up\ket +\beta | \down  \ket \rp_2 \\
        \left[ \lp \beta | \up\ket -\alpha | \down  \ket \rp \lp
        \beta^*
\bra \up |- \alpha^* \bra \down | \rp \right]_1  \lp \frac{
|\up\up\ket+|\down\down\ket }{\sqrt{2}}\rp_{12} = \frac{1}{\sqrt{2}}
\lp \beta | \up\ket -\alpha | \down  \ket \rp_1 \lp \beta^* | \up\ket -\alpha^* | \down  \ket \rp_2  \\
      \end{array}
    \right. \label{stateprep2}
 \ee

We interpret this as follows. Bob would like his particle's state to
be $\alpha | \up\ket +\beta | \down  \ket $ where $\alpha,~\beta$
are determined by Alice. She sets her measuring apparatus to measure
in the basis $\left\{\alpha^* | \up\ket +\beta^* | \down \ket,~\beta
| \up\ket -\alpha | \down \ket \right\} $. If she finds her particle
to be in the first state (which happens with probability 1/2), she
tells Bob that his particle has the desired state. If she doesn't,
she tells him it is in the orthogonal state. In the latter case, it
is clear from the form of the $\alpha,\beta$ dependence that Bob's
state is related to the desired state by an anti-unitary
transformation. We also see that we can understand the preparation,
when successful, in terms of the (sesqui-)linearity of the
projection operation. It should be noted that in an actual
experimental application of this scheme, the spinor picture
\emph{is} likely to play a role (e.g., in the settings of a
Stern-Gerlach apparatus).

\section{Quantum state teleportation and remote state preparation}

The last observation suggests a generalization. If the parameters
$\alpha,\beta$ are first encoded in the state of an auxiliary
particle, we can use a fixed basis for the measurements (now, joint
measurements of the auxiliary particle and Alice's particle).
Instead of the first line of Equation \ref{stateprep2} we write:

\be \left[\frac{1}{2} \lp |\up\up\ket+|\down\down\ket\rp \lp
\bra\up\up |+ \bra \down\down | \rp\right]_{12}\lp \alpha| \up \ket
+\beta|\down \ket \rp_1 \lp \frac{ |\up\up\ket+|\down\down\ket
}{\sqrt{2}}\rp_{23} = \frac{1}{2} \lp \frac{
|\up\up\ket+|\down\down\ket }{\sqrt{2}}\rp_{12} \lp \alpha| \up \ket
+\beta|\down \ket \rp_3 \label{tele1} \ee

Alice doesn't need to know the values of $\alpha,\beta$ to make the
measurements (the state of particle 1 can be prepared by someone
else). A related and very important difference from the previous
situation is that if particle 1 is originally in an entangled state
with another system, the effect of a measurement with this outcome
is to replace particle 1 by particle 3 in the overall state. This
can be verified by inspection--in the Dirac bracket notation, all we
have to do is let $\alpha,\beta$ take 'ket' rather than 'c-number'
values.

The two-particle state measurement is, of course, not determined by
Equation \ref{tele1}. Any measurement basis will determine a set of
linear transformations of the state of particle 1 into that of
particle 3. However, we would like them to be unitary (since this is
equivalent to the requirement that the inverse, corrective,
operations be unitary). The remarkable paper by Bennett et
al.\cite{bennet} uses the basis\cite{BMR}: $\left\{\frac{
|\up\up\ket\pm|\down\down\ket }{\sqrt{2}},\frac{
|\up\down\ket\pm|\down\up\ket }{\sqrt{2}}\right\}$ (this is known as
the Bell basis). Let us write out explicitly the rest of the
measurement outcomes\footnote{ Note that the sum of equations \ref{tele1}-\ref{tele1d} 
is $\lp \alpha| \up \ket +\beta |\down \ket \rp_1 |\psi^+\ket_{23}=
|\psi^+\ket_{12}\lp \alpha | \up \ket +\beta|\down \ket \rp_3 +
|\psi^-\ket_{12}\lp \alpha | \up \ket -\beta|\down \ket \rp_3 +
|\phi^+\ket_{12}\lp \alpha | \down \ket +\beta|\up \ket \rp_3 +
|\phi^-\ket_{12}\lp \alpha | \down \ket -\beta|\up \ket \rp_3$. 
This just Eq. 5 of \cite{bennet}, adapted to the case where the ``quantum channel'' consists of a pair
of spin-1/2 particles in a $\psi^+$, rather than $\phi^-$ state.
}:

\be \left[\frac{1}{2} \lp |\up\up\ket-|\down\down\ket\rp \lp
\bra\up\up |- \bra \down\down | \rp\right]_{12}\lp \alpha| \up \ket
+\beta|\down \ket \rp_1 \lp \frac{ |\up\up\ket+|\down\down\ket
}{\sqrt{2}}\rp_{23} = \frac{1}{2} \lp \frac{
|\up\up\ket-|\down\down\ket }{\sqrt{2}}\rp_{12} \lp \alpha| \up \ket
-\beta|\down \ket \rp_3 \label{tele1b}\ee

\be \left[\frac{1}{2} \lp |\up\down\ket+|\down\up\ket\rp \lp
\bra\up\down |+ \bra \down\up | \rp\right]_{12}\lp \alpha| \up \ket
+\beta|\down \ket \rp_1 \lp \frac{ |\up\up\ket+|\down\down\ket
}{\sqrt{2}}\rp_{23} = \frac{1}{2} \lp \frac{
|\up\down\ket+|\down\up\ket }{\sqrt{2}}\rp_{12} \lp \alpha| \down
\ket +\beta|\up \ket \rp_3 \label{tele1c}\ee

\be \left[\frac{1}{2} \lp |\up\down\ket-|\down\up\ket\rp \lp
\bra\up\down |- \bra \down\up | \rp\right]_{12}\lp \alpha| \up \ket
+\beta|\down \ket \rp_1 \lp \frac{ |\up\up\ket+|\down\down\ket
}{\sqrt{2}}\rp_{23} = \frac{1}{2} \lp \frac{
|\up\down\ket-|\down\up\ket }{\sqrt{2}}\rp_{12} \lp \alpha| \down
\ket -\beta|\up \ket \rp_3 \label{tele1d}\ee

And the correction operations Bob needs to perform in each case, are
readily seen to be unitary \newline (e.g.,
$|\up\ket\mapsto|\up\ket,~~|\down\ket \mapsto -|\down\ket$ for
(\ref{tele1b}) ).

Comment. Let $\left\{|u_i\ket_{12} \right\}_{i=1..4}$ be the basis
corresponding to any nondegenerate two particle measurement for our
two Pauli spinors. Then as noted by Schr\"{o}dinger, there exists
for $|u_i\ket$ a so-called Schmidt decomposition: \be
|u_j\ket=\sum_{i=1..2}\lambda^j_i|a^j_i\ket_1|b^j_i\ket_2 \ee where
$\left\{|a^j_i\ket \right\}_{i=1..2}$ and $\left\{|b^j_i\ket
\right\}_{i=1..2}$ are orthonormal bases of the respective single
particle Hilbert spaces. It is clear that for such a basis to induce
unitary transformations of the state of particle 1 to that of 3, for
all $i,j$ we should have
$|\lambda^j_i|=\frac{1}{\sqrt{2}}$\footnote{This is also the
condition for the $|u_i\ket$ to be maximally entangled states.}.
This suggests a direct generalization to the teleportation of states
in an arbitrary finite dimensional Hilbert space\footnote{In some
cases it might be advantageous to imbed the Hilbert space, by using
ancillary particles if necessary, into one with dimensionality
$2^k$, for some $k$, and teleport the state one 2-dimensional degree
of freedom ('qubit') at a time. This is possible by the remark below
Equation \ref{tele1}.}.

Finally, let us note that the Bell basis can be characterized  as
the simultaneous eigenstates of the operators
$\sigma_{z1}\sigma_{z2}$ (spin-z alignment) and
$\sigma_{x1}\sigma_{x2}$ (relative phase of the two terms). The
correction operators also have elegant expressions in the Pauli
algebra ($1, \sigma_{z}, \sigma_{x}, \sigma_{z}\sigma_{x}$;
respectively).

\section{Continuous Teleportation}

Consider now the teleportation of a particle with a one-dimensional
continuous degree of freedom. It is straightforward to write down
the formal analog of Equation \ref{tele1}:

\be \lp \int dx dx' |x,x\ket \bra x',x'|\rp_{1,2} \int dx''
\psi(x'') |x''\ket_1 \int dx'''|x''',x'''\ket_{2,3} = \int dx
|x,x\ket_{1,2} \int dx' \psi(x')|x'\ket_3 \ee

Thus, if Alice measures the state of the particles 1 and 2 in a
basis that includes the element $\int dx |x,x\ket$ and is fortunate
enough to find them in this particular state, she has successfully
'teleported' the state of particle 1 to particle 3. We now need
analogs for Equation \ref{tele1b}-\ref{tele1d}, i.e., to extend
$\int dx |x,x\ket$ to an appropriate basis of the two-particle
Hilbert space. We have considerable freedom. The analogy suggests
the following general form:

\be \int dx \exp(i\phi(x))|x,x'(x)\ket \ee

If we require that $x'(x)$ be smooth, the comment on the Schmidt
coefficients for the discrete case suggests that $\left|\frac{\dd
x'}{\dd x}\right| =1$, and so $x'(x)$ should have the form
$x'(x)=\pm x+\alpha$. The $\phi(x)$ should be chosen so as to ensure
completeness. The problem is to find a basis that corresponds to a
natural measurement.

Following the original continuous teleportation paper by Vaidman
\cite{mytele}, we note that the Bell basis, used in the discrete
case, can also be characterized as the eigenbasis of the operators
$\lp\sigma_{y1}+\sigma_{y2}\rp mod~ 4$ and
$\lp\sigma_{x1}+\sigma_{x2}\rp mod~ 4$ (which are linear in the
operators). These operators had been investigated much earlier by
Aharonov and Albert \cite{AA}, in the context of non-local
measurements. The analogous operators for the one-dimensional
continuous case were $x_1+x_2, p_1-p_2$ (as in the original EPR
paper). This led him to suggest both a new interpretation of quantum
teleportation, and a generalization to the continuous case. From the
present point of view, it suggests the eigenbasis of of the
operators $x_1-x_2, p_1+p_2$, namely, $\left\{\int dx \exp(-i\beta
x)|x,x+\alpha\ket \right\}_{\alpha,\beta\in\mathbb{R}}$(where it has
been tacitly assumed that $x$ values are given in terms of some
given unit making the expressions dimensionless). The analogs of
Eqns. \ref{tele1b}-\ref{tele1d} are then:

\begin{eqnarray} \lp \int dx dx' e^{i\beta(x-x')} |x,x+\alpha\ket \bra
x',x'+\alpha|\rp_{1,2} \int dx'' \psi(x'') |x''\ket_1 \int
dx'''|x''',x'''\ket_{2,3} = \nonumber \\
 \int dx e^{i\beta
x}|x,x+\alpha\ket_{1,2} \int dx' \psi(x')e^{-i\beta
x'}|x'+\alpha\ket_3~~\alpha,\beta\in\mathbb{R} \label{tele2}
\end{eqnarray}

And the (unitary) correction operations Bob needs to take when Alice
measures $x_1-x_2=-\alpha,~p_1+p_2=-\beta$ correspond to the
operations: $x_3\mapsto x_3-\alpha,~p_3\mapsto p_3-\beta$.

I shall not attempt to justify the formal formulas of this section
in terms of measure theory. That would defeat the aim to keep this
introduction elementary, but more importantly, the limiting case of
delta-function correlations considered here is not only highly
singular, but also physically unrealistic. The more important
question of realistic implementations was treated in the beautiful
paper of Braunstein and Kimble\cite{BK} which suggested a realizable
experiment in the realm of quantum optics. An elementary account of
their scheme is given in \cite{VER}.

\vspace{1 in}

I would like to thank Lev Vaidman for many helpful conversations on
teleportation. I gratefully acknowledge the support of the Welch
Foundation grant no. A-1261.


\end{document}